\documentclass[a4paper, 12pt]{article}

\usepackage{amsmath,amssymb,epsfig,subfigure}

\begin{document}

\numberwithin{equation}{section}

\title{Nonlinear transverse waves \\in deformed dispersive solids}
\author{Michel Destrade \& Giuseppe Saccomandi}

\date{}
\maketitle

\bigskip

\begin{abstract}

We present a phenomenological approach to dispersion in nonlinear
elasticity. A simple, thermomechanically sound, constitutive model
is proposed to describe the (non-dissipative) properties of a
hyperelastic dispersive solid, without recourse to a
microstructure or a special geometry. As a result, nonlinear and
dispersive waves can travel in the bulk of such solids, and
special waves emerge, some classic (periodic waves or pulse solitary
waves of infinite extend), some exotic (kink or pulse waves of
compact support). We show that for incompressible dispersive power-law
solids and forth-order elasticity solids, 
solitary waves can however only exist in the case of linear
transverse polarization.
We also study the influence of
pre-stretch and hardening. We provide links with other
(quasi-continuum, asymptotic) theories; 
in particular, an appropriate asymptotic multiscale expansion 
specializes our exact equations of motion to the vectorial 
MKdV equation, for any hyperelastic material.

\end{abstract}

\newpage

\section{Introduction}

The interplay of short and long range interactions in a physical phenomenon
leads to the existence of coherent structures that may\ play an important
role in the determination of the dynamical and thermodynamical properties of
real materials.
That is why highly sophisticated nonlinear \emph{lattice
models} taking account of long-range interactions have been extensively
used to investigate the complex behaviour of many materials, from
crystalline solids to rubber-like elastomers.
These models lead to major achievements in the study of those phenomena where inherent
characteristic lengths are a fundamental ingredient of the mechanical
behaviour, such as fracture mechanics or nano-structures (carbon nanotubes or
bio-polymers).

The mathematical models of atomic lattices involve large nonlinear systems
of ordinary differential equations and their treatment requires an enormous computational
effort.
Moreover, it is hard to conceive that such equations
may be amenable to a simple, yet detailed, mathematical analysis.
On the other hand, \emph{standard continuum theories}
-- which assume that the lattice parameter is equal
to zero -- are more appealing than their discrete counterparts
because they provide a powerful synthetic tool of analysis, description, and
prediction.
Their weakness is that they become all but useless at short wavelengths
(a well-known example of this flaw is recalled below.)
Hence the quest for \emph{quasi-continuum models}, which generalize
standard continuum mechanics to incorporate characteristic material
lengths.

A well-known example of quasi-continuum model is the so-called
\emph{strain-gradient elasticity theory}, which incorporates higher derivatives of
the displacement.
This theory is clearly connected with the higher terms in
the naive (Taylor) expansion in terms of small $h$ of a lattice.
To reveal this connection, let us consider the vibration of a single
1-D lattice, via the following Hamiltonian
\begin{equation}
H=\sum_{n}\left[ \overset{.}{y}_{n}^{2}(t)+P\left( \frac{y_{n}-y_{n-1}}{h}%
\right) \right] ,  \label{i1}
\end{equation}
where $y_{n}$ is the displacement from equilibrium of the $n-$th particle,
$h$ is the equilibrium inter-particle distance, and $P$ is the inter-particle
potential.
The equations of motion derived from \eqref{i1} are
\begin{equation}
u_{n, tt}=\left[ T(u_{n+1})+T(u_{n-1})-2T(u_{n})\right] /h^{2},  \label{i2}
\end{equation}
where  $u_{n}=(y_{n}-y_{n-1})/h$ and $T(\cdot )=P^{\prime }(\cdot )/h$.
The continuum limit of this equation, up to $O(h^{4})$, is
\begin{equation}
u_{tt}=\left[ T(u)\right] _{xx}+\frac{h^{2}}{12}u_{xxxx}.  \label{i4}
\end{equation}
Here it is the last term (fourth-derivative with respect to space) which
is usually hailed as being the first correction needed to account for
the dispersive effects due to discreteness.

Mindlin \cite{mindlin} introduced phenomenological strain-gradient
theories of continuum mechanics in the early 1960s; Green and
Rivlin \cite{green} extended them to include strain-gradients of
any order; and Toupin and Gazis \cite{toupin} established a
rigorous correspondence between strain-gradient theories and
atomic lattices. Equation \eqref{i4} presents an example of this
connection. This equation is actually quite popular in all those
applications of continuum mechanics where inherent characteristic
lengths of materials may not be neglected, even though it is
associated with several unpleasant problems. For instance, because
of the presence of a fourth-order space derivative term,
additional boundary conditions are required. Also, the
initial-value problem for \eqref{i4} is ill-posed. Indeed, the
associated dispersion relation is
\begin{equation}
\frac{\omega^2}{c^2} = K^2 - \frac{h^2}{12} K^4,  \label{i5}
\end{equation}
in the linear case, where $c$ is the speed of sound,
$\omega$ is the frequency, and $K$ is the wave number.
Clearly here, instability develops at short wavelengths
(when $K^2 h^2 > 12$).
Rosenau \cite{rosenau} overcame this problem by starting from the  exact
dispersion relation for the linear discrete system \eqref{i2}, that is
\begin{equation}
\frac{\omega^2}{c^2} =\frac{4}{h^2} \sin^2 \left(h K/2 \right),  \label{i6}
\end{equation}
to deduce a regularizing expansion in the parameter $h$,
\begin{equation}
\frac{\omega^2}{c^2} = \frac{K^2} {1 + h^2 K^2/12}.  \label{i7}
\end{equation}
It is clear that equations \eqref{i5} and \eqref{i7} are
equivalent at large wavelengths (small $K$),
but not at short wavelengths (large $K$),
where the latter remains bounded.
Now rewrite the continuum limit of \eqref{i2} as
\begin{equation}
u_{tt} = L_A D^2 \left[T(u) \right],
\quad \text{where} \quad
D \equiv \partial_{x},  \label{i8}
\quad
L_A \equiv 1 + \frac{h^2}{12} D^2 + \frac{h^4}{360} D^4 + \ldots  
\end{equation}
Then observe that $L_A$ is an invertible Schroedinger operator, with
\begin{equation}
L_A^{-1} = 1 - \frac{h^2}{12} D^2 + \frac{h^4}{240} D^4 + \ldots.  \label{i10}
\end{equation}
Applying the operator $L_A^{-1}$ to both sides of \eqref{i8} gives,
up to $O(h^{4})$, the equation
\begin{equation}
u_{tt} = \left[T(u) \right]_{xx} + \frac{h^2}{12} u_{xxtt}.  \label{i11}
\end{equation}
In the linear limit, this equation possesses the (bounded)
dispersion relation \eqref{i7}. Also, the Cauchy problem for this
equation is well-posed and we do not require additional boundary
conditions \cite{maugin}.

Rubin, Rosenau, and Gottlieb \cite{RRG} elaborated a phenomenological 3D
counterpart to the regularization procedure of Rosenau, by
modifying the Cauchy stress tensor $\mathbf{T}$ and the free energy $\psi$
to include dispersive effects,
while keeping the other thermomechanical entities (the entropy, the
internal production of entropy, and the entropy flux) unchanged.
Destrade and Saccomandi \cite{DS06a}, \cite{DS06} recently showed
that the material model of Rubin, Rosenau, and Gottlieb falls within
the theory of simple materials, which is a theory where the Cauchy stress
tensor depends only on the history of the
deformation gradient.
The consequence is that it is possible to introduce inherent characteristic
lengths without having to introduce exotic concepts such as the concept
of \emph{hyperstress}, necessary to deal with strain-gradient theories.

The aim of the present paper is to provide a general treatment
for the propagation of large amplitude transverse bulk waves in the framework
of a phenomenological theory.
Hence we aim at generalizing the celebrated study of Gorbacheva and Ostrovsky
\cite{Gorba}, which presented results for the continuum
limit of a one-dimensional lattice with elastic bonds under longitudinal
stress.
Their results were obtained only for a special law of the elastic bond and
for a single progressive wave
(in a continuum KdV limit of the lattice).
Here we consider a \emph{general constitutive law}
for the strain-energy of the material and we also consider the
possibility of \emph{nonlinear dispersion}.
Thus we expect to be able to extend the results of
Gorbacheva and Ostrovsky to Murnaghan's materials, to Ogden's materials, and to
many other popular phenomenological response functions.
Focusing on incompressible solids,
and using the methods developed by Destrade and Saccomandi
\cite{DS05} \cite{DS06}, we present the governing equations in Section 2 and 
derive in Section 3 a single complex wave equation for
finite-amplitude transverse principal waves,
valid for any dispersive solid.
In Section 4 we specialize the equation to the case of a strain-hardening
power-law solid, and discuss the influence of polarization, dispersion,
pre-stretch, and strain-hardening.
We find some exotic solutions such as pulses or kinks with infinite or even
compact support.
However the analysis soon reveals that these localized solutions are the
exception rather than the rule.
The analysis is exact and does not rely on regularized expansions.
When we do perform such an expansion (Section 5) we
recover, within an appropriate asymptotic
limit, a ``vector Modified Korteveg-deVries equation'' (MKdV).

\section{Governing equations}

Let the motion of a body be described by
$\mathbf{x}=\mathbf{x}(\mathbf{X},t)$,
where $\mathbf{x}$ denotes the current coordinates of a point occupied
at time $t$  by
the particle which was at $\mathbf{X}$ in the reference configuration.
The associated \emph{deformation gradient} is
$\mathbf{F}(\mathbf{X},t) \equiv \partial \mathbf{x} / \partial \mathbf{X}$
and the \emph{spatial velocity gradient} is
$\mathbf{L}(\mathbf{X},t) \equiv \partial \mathbf{v}/\partial \mathbf{x}$,
where $\mathbf{v} = \partial \mathbf{x}/\partial t$ is the velocity vector.
We consider a material whose mechanical behaviour is described by a
given Cauchy stress tensor, $\mathbf{T}^{\text{E}}$ say.
Standard continuum theories (such as the linear theory of elasticity or the
Navier-Stokes equations of fluid mechanics) are intrinsically
size-independent;
to overcome this shortcoming we modify the
standard Cauchy stress tensor $\mathbf{T}^{\text{E}}$ so that the equation
for the full Cauchy stress tensor is
\begin{equation}
\mathbf{T}=\mathbf{T}^{\text{E}}+\mathbf{T}^{\text{D}},  \label{b2}
\end{equation}
where the new stress tensor term, $\mathbf{T}^{\text{D}}$,
is introduced to take into account dispersive effects.
Guided by preliminary work \cite{DS05}, \cite{DS06},  we take it in the form
\begin{equation}
\mathbf{T}^{\text{D}} = \alpha(\mathbf{D}\cdot \mathbf{D})
 [\mathbf{A}_{2} - \mathbf{A}_{1}^{2}],
\label{b3}
\end{equation}
where $\mathbf{D}$ is the stretching tensor,
$\mathbf{A}_{1}$ and $\mathbf{A}_{2}$ are the first two
Rivlin-Ericksen tensors,
\begin{equation}
 \mathbf{D} \equiv (\mathbf{L}+\mathbf{L}^{\text{T}})/2,
\qquad \mathbf{A}_{1} \equiv 2\mathbf{D},
 \qquad
\mathbf{A}_{2} \equiv \mathbf{\dot{A}}_{1} + \mathbf{A}_{1}\mathbf{L}
   + \mathbf{L}^{\text{T}}\mathbf{A}_{1},  \label{b4}
\end{equation}
and the \emph{dispersion material function}
$\alpha = \alpha (\mathbf{D}\cdot \mathbf{D})$,
must be positive due to thermodynamics restrictions.

It turns out that the term (\ref{b3}) is exactly the
one proposed by Rubin et al. in \cite{RRG},
and that $\mathbf{T}^{\text{D}}$ is a special case of the
extra Cauchy stress tensor
associated with a non-Newtonian fluid of second grade,
which is
\begin{equation}
\nu \mathbf{A}_{1}+\alpha _{1}\mathbf{A}_{2}+\alpha _{2}\mathbf{A}_{1}^{2},
\label{b7}
\end{equation}
in general, where $\nu $ is the classical viscosity and $\alpha _{1},\alpha _{2}$ are
the microstructural coefficients.
Note that \eqref{b3} is of the same form as \eqref{b7}
when $\nu =0$ and $\alpha_{1} + \alpha_{2} = 0$;
here the first equality means that the material is \emph{non-dissipative}, and
the second makes the model compatible with the laws of thermodynamics,
see Fosdick and Yu \cite{fosdick} for details
(If we let $\nu \neq 0$, then we have
the possibility to include dissipation as in the classical
Navier-Stokes theory, but we do not pursue that alley here.)
The coincidence between \eqref{b3} and the model of Rubin et al.
is completed once we identify the dispersion material function $\alpha$
with the derivative of the Helmholz free energy $\psi$ which they introduced
to model dispersion:
\begin{equation} \label{alpha_psi}
 \alpha
  = \psi' (\mathbf{D}\cdot \mathbf{D}).
\end{equation}

Now we restrict our attention to homogeneous, isotropic, compressible
elastic solids.
The response of such materials from an undeformed
reference configuration is described by the constitutive relation
\begin{equation}
\mathbf{T}^{\text{E}} =
\left(\dfrac{I_2}{\sqrt{I_3}} \dfrac{\partial \Sigma}{\partial I_2}
 + \sqrt{I_3} \dfrac{\partial \Sigma}{\partial I_3} \right) \mathbf{I}
  + \dfrac{2}{\sqrt{I_{3}}} \dfrac{\partial \Sigma}{\partial I_1}
      \mathbf{B}
     + 2\sqrt{I_{3}}  \dfrac{\partial \Sigma}{\partial I_2}\mathbf{B}^{-1},
     \label{b8}
\end{equation}
where $\mathbf{I}$ is the identity tensor, $\mathbf{B}$ is the left Cauchy-Green
strain tensor defined by $\mathbf{B}=\mathbf{FF}^{\text{T}},$ and the
strain-energy function $\Sigma $ is a function of the first three invariants
$I_1$, $I_2$, $I_3$ of $\mathbf{B}$, defined in turn by
\begin{equation}
I_{1}=\text{tr }\mathbf{B},\qquad
 I_{3}=\text{det } \mathbf{B}, \qquad
I_{2}= \textstyle{\frac{1}{2}}
 \left[I_1^{2} - \text{tr }(\mathbf{B}^{2})\right]
  = I_{3} \text{tr }(\mathbf{B}^{-1}).  \label{b9}
\end{equation}

When the material is \emph{incompressible}, $\text{det } \mathbf{F} =1$
at all times (every motion is isochoric), and we replace \eqref{b8} with
\begin{equation} \label{elastic}
\mathbf{T}^{\text{E}} =
  -p\mathbf{I} + 2\Sigma_1 \mathbf{B} + 2\Sigma_2 \mathbf{B}^{-1},
\end{equation}
where now $\Sigma = \Sigma(I_1, I_2)$ only (because $I_3 = 1$ at all times),
$p$ is the Lagrange multiplier due to the incompressibility constraint,
and  $ \Sigma_i \equiv \partial \Sigma / \partial I_i$.

For the rest of the paper, we focus on incompressible solids and we
try to work in all generality for the
elastic part of the Cauchy stress tensor.
For applications and illustrative examples, we shall consider
for instance the \emph{power-law model}
proposed by Knowles \cite{Kn},
\begin{equation}
\Sigma =\frac{\mu }{2b}\left[ \left( 1+\frac{b}{n}\left( I_{1}-3\right)
\right) ^{n}-1\right],  \label{11}
\end{equation}
where $\mu>0$ is the infinitesimal shear modulus,
and $b >0$, $n >0$ are material constants.
The neo-Hookean model, for which the generalized shear modulus is constant,
corresponds to the case $n=1$.
The material modelled by \eqref{11} is \emph{softening} in simple shear
when $n<1$ and it is \emph{hardening} in simple shear
when $n>1$;
hence Raghavan and Vorp \cite{vorp} recently modelled
the elasticity of aortic abdominal aneurysms by taking $n = 2$;
also, on letting $n \rightarrow \infty$ we find Fung's exponential strain-energy density,
widely used to describe soft biological tissues.
We also mention models which are popular in the so-called weakly
nonlinear elasticity theory, based on truncated polynomial
expansions of the strain energy density.
For instance, finite element packages often propose a
\emph{polynomial strain energy potential};
the second degree version of this potential is
\begin{equation} \label{polynomial}
\Sigma = C_{10}(I_1-3) + C_{01}(I_2-3) +  C_{20}(I_1-3)^2
  + C_{02}(I_2-3)^2 + C_{11}(I_1-3) (I_2-3),
\end{equation}
where the $C_{ij}$ are five material parameters to be adjusted for
optimal curve fitting. Note that this strain energy density
coincides with the power-law $n=2$ model when $I_1 = I_2$. 
Another important model comes from the nonlinear acoustic literature,
where the strain energy function is expanded up to the fourth order in the
strain in order to reveal nonlinear shear waves. In that
framework, Murnaghan's expansion \cite{murn} is often used:
\begin{equation}
\Sigma =\dfrac{\lambda +2\mu }{2}i_{1}^{2}-2\mu i_{2}+\dfrac{l+2m}{3}%
i_{1}^{3}-2mi_{1}i_{2}+ni_{3}+\nu _{1}i_{1}^{4}+\nu
_{2}i_{1}^{2}i_{2}+\nu _{3}i_{1}i_{3}+\nu _{4}i_{2}^{2},
\label{4thOrder}
\end{equation}
where $\lambda $, $\mu $ are the Lam\'{e} moduli, $l$, $m$, $n$ are
the third-order moduli, and $\nu _{1}$, $\nu _{2}$, $\nu _{3}$,
$\nu _{4}$ are the fourth-order moduli. 
Here we used
the first three principal invariants $i_{1}$, $i_{2}$, $i_{3}$ of
$\mathbf{E}$, the Green-Lagrange
strain tensor; they are related to the first three principal invariants $%
I_{1}$, $I_{2}$, $I_{3}$ by the relations,
\begin{equation}
I_{1}=2i_{1}+3,\quad I_{2}=4i_{1}+4i_{2}+3,\quad
I_{3}=2i_{1}+4i_{2}+8i_{3}+1.  \label{invariants4}
\end{equation}
When the material is incompressible, i.e. the relative portion of
energy stored in compression is negligible, Hamilton et al.
\cite{HZ} proposed the following reduced version of the expansion
\eqref{4thOrder},
\begin{equation}  \label{hamil}
\Sigma = -2 \mu i_2 + n i_3 + \nu_4 i_2^2.
\end{equation}

Similarly,  in the examples we specialize the \emph{dispersion material
function} of \eqref{b3}  and \eqref{alpha_psi} to the following simple form,
\begin{equation}
\alpha = \rho \beta_0 + \rho \beta_1 (\mathbf{D} \cdot \mathbf{D}),
\quad \text{so that} \quad
\psi = \rho \beta_0 (\mathbf{D} \cdot \mathbf{D}) + \rho \frac{\beta_1}{2} (\mathbf{D} \cdot \mathbf{D})^2,
\label{13}
\end{equation}
where $\beta_0$ and $\beta_1$ are positive material constants.

\textbf{Remark 1}:
We emphasize that although the stretching tensor $\mathbf{D}$ and its objective
time derivative appear explicitly in the expression \eqref{b3},
the dispersive stress tensor $\mathbf{T}^{\text{D}}$ nonetheless does not
contribute to dissipation.
In that respect, its action is analogue to a gyroscopic force term in
classical mechanics.
\\

\textbf{Remark 2}: That $\mathbf{T}^{\text{D}}$ gives a constitutive
equation compatible with the class of simple materials is a
happenstance.
If we were to consider further terms in the approximation of the
discrete linear dispersion relation \eqref{i6}, this property would no longer
be true.
Hence, to approximate \eqref{i6} to order $O(h^{6})$,
it is necessary to introduce the gradient of the strain. Then, the
quasi-continuum modelling of the material has to turn to more complex theories
and the mathematical simplicity and feasibility of the present model is lost.
\\

\textbf{Remark 3}: We note that the regularized equation  \eqref{i11} of Rosenau,
and the constitutive model \eqref{b2} both involve derivatives with respect to time,
and are thus mainly suitable to study weak-nonlocality in the
framework of \emph{wave propagation}.
\\

\textbf{Remark 4}: We focus mostly on incompressible solids 
because it is a convenient way to bypass the problem of 
a coupling between transverse and longitudinal waves, 
and to build a general three-dimensional generalization 
of the one-dimensional lattice theory used by Gorbatcheva and Ostrosky
\cite{Gorba}, where transverse waves were
considered coupled to a longitudinal static deformation
(note that Cadet \cite{cadet}, \cite{cadet87b} considered a lattice model
with  longitudinal/transverse wave coupling.) 
In compressible materials,  transverse waves are in general coupled to the
longitudinal wave, except for those special forms of the strain energy
density function which satisfy $\Sigma_1 + 2 \Sigma_2 + \Sigma_3 = \text{const.}$,
see the survey \cite{sacco06} for details. 
The fourth-order elasticity model \eqref{hamil} clearly fits into 
that category.

\section{Finite amplitude waves in incompressible solids}

In this Section we consider the following class of motions
\begin{equation}
x=\lambda^{-\textstyle{\frac{1}{2}}}  X+u(z,t),
\qquad
y= \lambda^{-\textstyle{\frac{1}{2}}} Y+v(z,t),
\qquad
z=\lambda Z,  \label{pw1}
\end{equation}
where $\lambda$ is the pre-stretch in the $Z$ direction.
These are motions describing a transverse wave, polarized in the ($XY$)
plane, and propagating in the $Z$ direction of a solid subject to a pure
homogeneous equi-biaxial pre-stretch along the $X$, $Y$, and $Z$ axes, with
corresponding constant principal stretch ratios
$\lambda^{-\textstyle{\frac{1}{2}}} $,
$\lambda^{-\textstyle{\frac{1}{2}}} $,
and $\lambda$, respectively.
Here $u$ and $v$ are yet unknown scalar functions of $z$ and $t$.

\subsection{Equation of motion}

The geometrical quantities of interest are the left
Cauchy-Green strain tensor $\mathbf{B}$ and its inverse $\mathbf{B}^{-1}$,
given by
\begin{equation}
\begin{bmatrix}
\lambda^{-1} + \lambda^2 u_z^2 & \lambda^2 u_z v_z  & \lambda^2 u_z \\
\lambda^2 u_z v_z & \lambda^{-1} + \lambda^2 v_z^2 & \lambda^2 v_z \\
\lambda^2 u_z & \lambda^2 v_z & \lambda^2
\end{bmatrix},  \label{pw3bis}
\qquad
\begin{bmatrix}
\lambda & 0 & -\lambda u_z \\
0 & \lambda & -\lambda v_z \\
-\lambda u_z & -\lambda v_z & \lambda (u_z^2 + v_z^2) + \lambda^{-2}
\end{bmatrix},
\end{equation}
respectively, where the subscript denotes partial differentiation.
The first two invariants of strain are thus
\begin{equation}
I_1 = 2 \lambda^{-1} + \lambda^2 + \lambda^2 (u_z^2 + v_z^2),
\quad
I_2 = 2 \lambda + \lambda^{-2} + \lambda (u_z^2 + v_z^2),  \label{pw6}
\end{equation}
and they are clearly functions of $u_z^2 + v_z^2$ only, which itself
depends on $z$ and $t$ only.
The kinematical quantities of interest are $\mathbf{A}_{1}$, $\mathbf{A}_{1}^2$,
and $\mathbf{A}_{2}$, given by
\begin{equation}
\begin{bmatrix}
0 & 0 & u_{zt} \\
0 & 0 & v_{zt} \\
u_{zt} & v_{zt} & 0
\end{bmatrix},
 \quad
\begin{bmatrix}
u_{zt}^2 & u_{zt} v_{zt} & 0 \\
u_{zt} v_{zt} & v_{zt}^2 & 0 \\
0 & 0 & u_{zt}^{2}+v_{zt}^{2}
\end{bmatrix},
 \quad
\begin{bmatrix}
0 & 0 & u_{ztt} \\
0 & 0 & v_{ztt} \\
u_{ztt} & v_{ztt} & 2(u_{zt}^{2}+v_{zt}^{2})
\end{bmatrix},  \label{pw5}
\end{equation}
respectively.
Hence we find that $\mathbf{D \cdot D}$, and thus $\alpha$ and $\psi$, are
functions of $u_{zt}^{2}+v_{zt}^{2}$ only:
\begin{equation} \label{DD}
 \mathbf{D \cdot D} = \textstyle{\frac{1}{2}} (u_{zt}^{2}+v_{zt}^{2}).
\end{equation}

Now the equations of motion, in the absence of body forces, are given in
current form as: $\text{div }\mathbf{T}=\rho \partial^2 \mathbf{x} / \partial t^2$,
where $\rho$ is the mass density.
For an \emph{incompressible} material, they read
\begin{equation}
-\dfrac{\partial p}{\partial x} + \dfrac{\partial T_{13}}{\partial z}
 = \rho u_{tt}, \qquad
-\dfrac{\partial p}{\partial y} + \dfrac{\partial T_{23}}{\partial z}
 = \rho v_{tt},\qquad
\dfrac{\partial T_{33}}{\partial z} = 0.
\label{pw7}
\end{equation}

Differentiating these equations with respect to $x$, we find
$p_{xx} = p_{yx} = p_{zx} = 0$,
so that $p_{x} = q_1(t)$, say.
Similarly, by differentiating the equations with respect to $y$,
we find $p_{y} = q_{2}(t)$, say.
Now the first two equations reduce to
\begin{equation}
 -q_{1}(t)+(Q u_{z})_{z} + \left(\alpha u_{z t t} \right)_{z}
 = \rho u_{t t},
\qquad
 -q_{2}(t)+(Q v_{z})_{z}+\left( \alpha v_{z t t}\right) _{z}=\rho v_{t t},
\label{pw9}
\end{equation}
and the third equation determines $p$. Here, $Q=Q(u_{z}^{2}+v_{z}^{2})$ is
the \emph{generalized shear modulus}, defined by
\begin{equation} \label{Q}
Q=2(\lambda ^{2} \Sigma_1 + \lambda \Sigma_{2}),
\end{equation}
and $\alpha  = \alpha (u_{z t}^{2}+v_{z t}^{2})$ is the material
function describing dispersive effects. 
From the generalized shear
modulus, it is possible to derive the shear stress-shear strain law
for the motion of an incompressible material. Strong experimental
evidence --- at least for rubber-like materials --- suggests that 
$Q>0$ for any isochoric deformation \cite{Beatty}.

Following Destrade and Saccomandi \cite{DS05}, we take the derivative of
Eqs.~(\ref{pw9}) with respect $z$, we introduce the \emph{strains} $U \equiv u_{z}$,
$V \equiv v_{z}$ and the complex function $W \equiv U+iV$, and we recast Eq.~(\ref{pw9})
as the following single complex equation,
\begin{equation}
(QW)_{zz} + \left(\alpha W_{tt} \right)_{zz} = \rho W_{tt},  \label{pw10}
\end{equation}
where $Q$ is now a function of $U^{2}+V^{2}$ alone:
$Q=Q(U^{2}+V^{2})$, and $\alpha$ is now a function of
$U_{t}^{2}+V_{t}^{2}$ alone: $\alpha =\alpha
(U_{t}^{2}+V_{t}^{2})$.

It is convenient to decompose the complex function $W$ into its modulus $\Omega$ (say)
and its argument $\theta$ (say),
\begin{equation}
W (z, t) = \Omega(z,t) \text{e}^{\text{i} \theta (z,t)},
\quad \text{so that} \quad
Q = Q(\Omega^2), \quad
\alpha = \alpha(\Omega_t^2 + \Omega^2 \theta_t^2).
\label{pw11}
\end{equation}
Here $\Omega$ is the \emph{strain wave amplitude} and $\theta$ is the
\emph{angle of polarization}.
Note that the combinations of \eqref{Q} and \eqref{pw6} and of
\eqref{alpha_psi} and \eqref{DD} give the relations:
\begin{equation} \label{antiderivatives}
Q = 2 \frac{\text{d} \Sigma}{\text{d} (\Omega^2)},
\qquad
\alpha = 2 \frac{\text{d} \psi}{\text{d} (\Omega_t^2 + \Omega^2 \theta_t^2)}.
\end{equation}



\subsection{Travelling transverse waves}

Now we focus on travelling wave solutions in the form
\begin{equation}
W = W(s) = \Omega(s)\text{e}^{\text{i} \theta(s)},
\quad \text{where} \quad
s = z - c t,  \label{tw}
\end{equation}
$c$ being the speed.
This ansatz reduces (\ref{pw10}) to
$(Q W)'' + c^{2}(\alpha  W '')'' = \rho c^2 W''$.
We integrate it twice, taking  each integration constant
to be zero in order to eliminate the
rigid and the homogeneous motions.
We end up with a vector nonlinear oscillator,
\begin{equation}
\left(Q - \rho c^{2}\right) W + c^{2} \alpha W'' = 0.  \label{tw1}
\end{equation}
Then, separating the real part of this equation from the imaginary part
gives
\begin{align}
& \left[Q(\Omega^2) - \rho c^2 \right] \Omega
    +\alpha\left(c^2 \Omega^{'2} + c^2 \Omega^2 \theta^{' 2}\right)
      c^2 (\Omega'' - \theta^{' 2} \Omega) = 0,
\notag \\
& \left(\Omega^2 \theta'\right)'=0.  \label{tw3}
\end{align}
We integrate the latter equation to 
$\theta' = I \Omega^{-2}$, 
where $I$ is a constant, and we substitute into the former equation to get
\begin{equation} \label{governing}
\left[Q(\Omega^2) - \rho c^2\right] \Omega
  +\alpha\left(c^2 \Omega^{' 2} + c^2 I^2 \Omega^{-2}\right)
       c^2(\Omega'' - I^2 \Omega^{-3}) =0.
\end{equation}
Multiplying across by $\Omega'$, and using the connections (\ref{antiderivatives}),
we find the \emph{energy first integral},
\begin{equation} \label{1st_integral}
 \Sigma(\Omega^2) -\textstyle{\frac{1}{2}} \rho c^2 \Omega^2
   + \psi\left(c^2 \Omega^{' 2} + c^2 I^2 \Omega^{-2}\right)  = E,
\end{equation}
where $E$ is a constant.

Now a standard analysis using phase-plane theory \cite{Gorba} or a
Weierstrass ``potential'' theory \cite{peyrard}, \cite{sacco04}
gives a classification of  all possible travelling waves
solutions. Note that the choice $I=0$ gives $\theta =$ const., 
and then we obtain the special case of a
linearly-polarized transverse wave.

Eq.\eqref{1st_integral} is the central equation of this paper.
It is valid for any dispersive incompressible solid with constitutive
equations \eqref{b2}, \eqref{b3}, \eqref{alpha_psi}, and \eqref{elastic}.
It is exact and no approximations nor regular expansions have been made to
derive it.
Before we move on to a constitutive example, we check that the equation is coherent
with quasi-continuum theories.

\subsection{Link with nonlinear lattice theory}

The constitutive model \eqref{b2}, \eqref{b3}, \eqref{elastic} is grounded
on quasi-continuum models,
and we expect that  \eqref{1st_integral} covers atomistic theories.
In fact the correlation is remarkably simple:
it suffices to take $\alpha =$ const. (i.e. to take $\beta_1 = 0$ in \eqref{13})
and to take $Q = \mu_0 + \mu_1 \Omega^2$ where $\mu_0$, $\mu_1$ are positive
constants (i.e. $\Sigma$ is of the polynomial form (\ref{polynomial}),
see \cite{DS06}).
Then Eq.\eqref{governing} reduces to
\begin{equation}
\Omega'' + \left[
\frac{\mu_1}{\rho c^2 \beta_0}\Omega^2
 - \frac{1}{\beta_0}\left(1 - \frac{\mu_0}{\rho c^2}\right)
 - \frac{I^2}{\Omega^4} \right]\Omega = 0.
 \label{gorba_eq}
\end{equation}
Gorbatcheva and Ostrosky \cite{Gorba} and Cadet \cite{cadet} 
obtained this equation for a mono-atomic chain
with interatomic interactions by a power series
expansion up to the fourth order over the lattice parameter.

\section{Travelling waves in hardening power-law solids}

To investigate the effect of nonlinearity in the elasticity and in the dispersion,
we first take a look at the situation for
transverse travelling waves in dispersive hardening power-law solids,
for which $\Sigma$ is given by (\ref{11}) and $\psi$ by (\ref{13}).
We start with the material at $n=2$, and we indicate the trends for $n>2$.

\subsection{Linearly polarized transverse waves}

When $n=2$ and the material is not pre-stretched ($\lambda=1$), we find that 
\begin{equation} \label{eqn_n_2}
\Sigma(\Omega^2) = \frac{\mu}{2} \left( 1 + \frac{b}{4} \Omega^2 \right)\Omega^2.
\end{equation}
Then, the motion of
linearly-polarized waves ($I=0$) is governed by the
following specialization of \eqref{1st_integral},
\begin{equation} \label{eqn_pw}
 (\mu - \rho c^2 + \frac{\mu b}{4}\Omega^2)\Omega^2
  + \rho c^2(\beta_0 \Omega^{'2}  + \frac{\beta_1}{4} \Omega^{'4}) = 2 E.
\end{equation}

First we investigate the situation when $\beta_0 \ne 0$, $\beta_1 = 0$.
Then we recover the classic solutions of quasi-continuum modelling of nonlinear
lattices.
To show this, we call $\Omega_0$ the value of $\Omega$ when $\Omega^{'2} =0$,
and we favour $\Omega_0$ in stead of $E$ as an arbitrary constant
(they are related through $2 E =(\mu - \rho c^2)\Omega_0^2 + \mu b \Omega_0^4/4$).
Then we perform the following change of variable and change of function,
\begin{equation} \label{changes1}
 \xi =  \sqrt{\frac{\mu b \Omega_0^2}{4 \rho c^2 \beta_0}}s,
 \qquad
 \omega(\xi) \equiv \frac{\Omega(s)}{\Omega_0},
\end{equation}
and the non-dimensional version of \eqref{1st_integral} at $\beta_1 = 0$ is
\begin{equation}
\omega^{' 2} = -(\omega^2 - 1)\left[\omega^2 + 1 - \frac{4}{\mu b \Omega_0^2}(\rho c^2 - \mu)\right].
  \label{non_dimensional1}
\end{equation}
For the choice $E=0$ (or equivalently $\Omega' = 0$ at $\Omega = 0$),
the changes of variable and of function are
\begin{equation}\label{changes2}
 \xi =  \sqrt{\frac{\rho c^2 - \mu}{\rho c^2 \beta_0}}s,
 \qquad
 \omega(\xi) \equiv \sqrt{\frac{\mu b}{4(\rho c^2 - \mu)}}\Omega(s),
\end{equation}
and the non-dimensional equation is
\begin{equation}
\omega^{' 2} = - \omega^2 (\omega^2 - 1).
 \label{non_dimensional2}
\end{equation}
In the latter case, the wave speed is necessarily supersonic with respect to an
infinitesimal bulk wave (i.e. $\rho c^2 > \mu$) but otherwise arbitrary,
and the solution is a well-known \emph{pulse solitary wave},
$ \omega (\xi) = \text{ sech } \xi$, shown on Figure 1(a).
\begin{figure}
\centering
\epsfig{figure=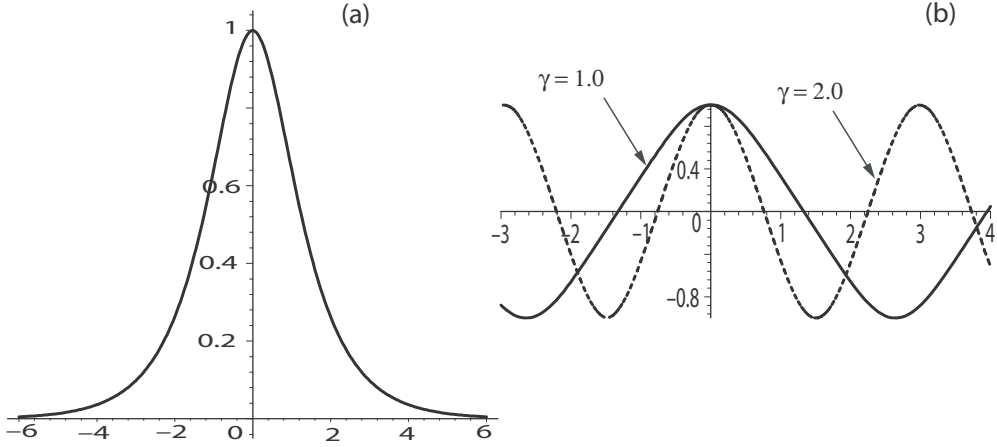, width=.95\textwidth}
  \caption{Nonlinear waves in a power-law dispersive solid: (a) pulse, (b) periodic.}
\end{figure}
In the former case \eqref{non_dimensional1}, the solutions are \emph{periodic waves} in general.
When the arbitrary speed is such that $\rho c^2 < \mu + \mu b \Omega_0^2 / 4$,
the amplitude varies between -1 and +1 and the solution is found explicitely in terms of 
a Jacobi elliptic function as
\begin{equation}
\omega (\xi) = \text{cn} \left(\sqrt{1 - \gamma^2} \xi \; | \; 1/\sqrt{1 - \gamma^2} \right),  
\quad \text{where } 
 \gamma = \sqrt{1 + 4(\mu - \rho c^2)/(\mu b \Omega_0^2)}.
\end{equation}
Notice that here the finite amplitude periodic wave can travel at the
same speed as that of an infinitesimal shear bulk wave $\sqrt{\mu/\rho}$.
Figure 1(b) displays this wave in solid curve when $\gamma = 1$
(and then $\rho c^2 = \mu$ and the wavelength is $\simeq 5.24$)
and in dashed curve when $\gamma = 2$
(and then the wavelength is $\simeq 2.97$).
When the speed is such that $\rho c^2 > \mu + \mu b \Omega_0^2 / 4$,
there are two possible waves, one with amplitude varying between
$-\sqrt{4(\rho c^2 - \mu)/(\mu b \Omega_0^2)-1}$ and -1, the other with amplitude varying between
$\sqrt{4(\rho c^2 - \mu)/(\mu b \Omega_0^2)-1}$ and +1.
Finally, there is a special wave for the choice $\rho c^2 = \mu + \mu b \Omega_0^2 / 4$,
because then Eq.\eqref{non_dimensional1} has the \emph{pulse solitary wave}
solution: $\omega =  \text{ sech } \xi$.

A different picture emerges when $\beta_0 = 0$, $\beta_1 \ne 0$.
In the case  $E \ne 0$, the changes of variable and of function,
\begin{equation}
 \xi =  \left[ \frac{\mu b }{\rho c^2 \beta_1} \right]^{1/4} s,
 \qquad
 \omega(\xi) \equiv \frac{\Omega(s)}{\Omega_0},
\end{equation}
give the non-dimensional governing equation
\begin{equation}
\omega^{' 4} = -(\omega^2 - 1)\left[\omega^2 + 1 - \frac{4}{\mu b \Omega_0^2}(\rho c^2 - \mu)\right],
  \label{non_dimensional3}
\end{equation}
whereas the choice $E=0$ and the changes
\begin{equation}
 \xi =  \left[{\frac{\mu b}{\rho c^2 \beta_1}} \right]^{1/4} s,
 \qquad
 \omega(\xi) \equiv \left[\frac{\mu b}{4(\rho c^2 - \mu)} \right]^{1/2} \Omega(s),
\end{equation}
give
\begin{equation}
\omega^{' 4} = - \omega^2 (\omega^2 - 1).
 \label{non_dimensional4}
\end{equation}
For  \eqref{non_dimensional3}, the solutions are \emph{periodic waves} in general,
just as when $\beta_0 \ne 0$, $\beta_1 = 0$, see above.
Equation  \eqref{non_dimensional4} however has  supersonic
\emph{solitary wave solutions with compact support},
a rare occurrence for waves in solids.
Indeed,  \eqref{non_dimensional4} can be integrated formally to give \cite{DS06},
$ \omega(\xi) =
   \mathcal{I}^{-1} (\xi)$,
where $\mathcal{I}^{-1}$ is
the inverse of a monotonic function defined in terms
of a hypergeometric function as
$\mathcal{I}(x) =  2(x^2)^{1/4} \text{Hyp}_2\text{F}_1[1/4, 1/4, 5/4, x^2]$.
Figure 2 shows how we can construct (weak) solutions with \textit{finite} support measures
as a \emph{compact solitary kink wave} or as a
\emph{compact solitary pulse wave}.
Notice also that these solutions also occurs for  \eqref{non_dimensional3}
at the special speed given by $\rho c^2 = \mu + \mu b\Omega_0^2/4$.
\begin{figure}
 \centering
  \mbox{\subfigure[kink]{\epsfig{figure=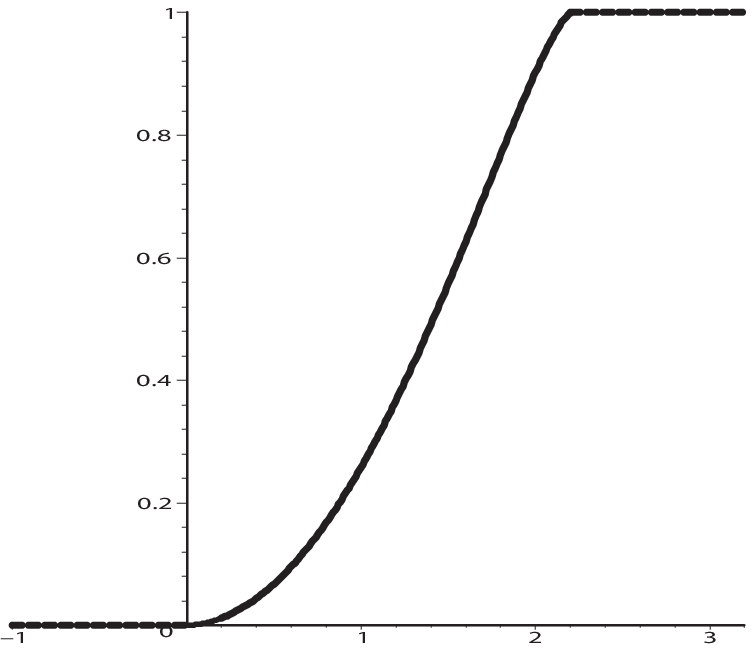, width=.45\textwidth, height=.35\textwidth}}
  \quad \quad
     \subfigure[pulse]{\epsfig{figure=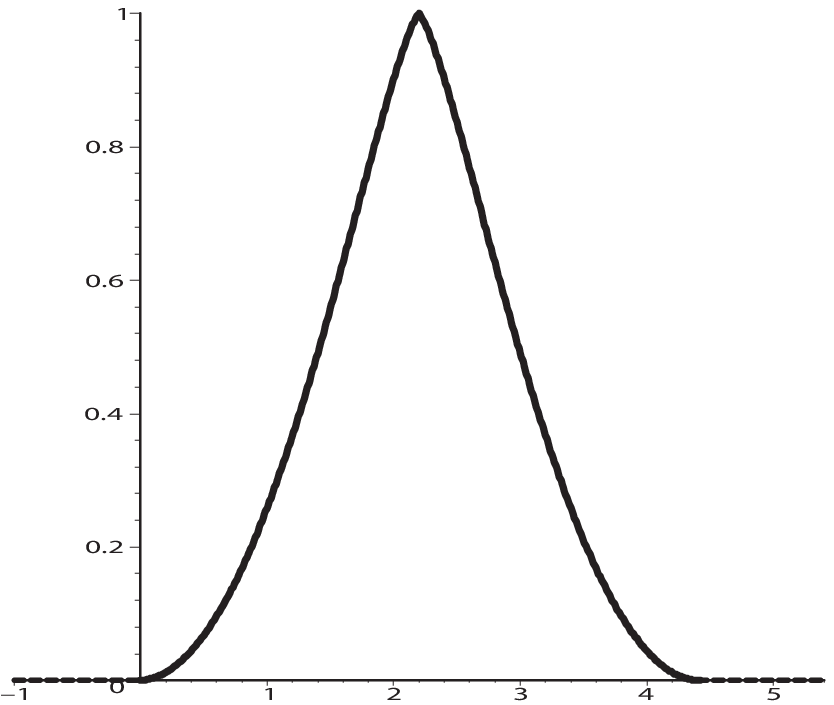, width=.45\textwidth, height=.35\textwidth}}}
\caption{Compact solitary waves in a power-law dispersive solid.
For the kink, the solution $\omega(\xi)$ is defined on the compact
support $[0, \pi/4^{1/4}]$; for the pulse, on $[0, 2\pi/4^{1/4}]$.}
\end{figure}

When $\beta_0 \ne 0$, $\beta_1 \ne 0$, the governing equation is a
quadratic in $(\Omega')^2$. 
Its resolution is straightforward and
the results are similar \cite{sacco04} to those in the case where
$\beta_0 \ne 0$, $\beta_1 = 0$ that is, solitary and periodic
 waves on infinite support measures (no compact waves). In
this case we checked, using the same arguments as those used in
\cite{sgura}, that when the ratio $\beta_0/\beta_1$ approaches zero
the tails of the non-compact localized wave decay more and more
rapidly and the solution approaches more and more the limiting
compact wave (obtained at $\beta_0=0$).

\subsection{Plane polarized transverse waves}

Now what happens when the wave is not linearly polarized ($I\neq 0$)?
Then we find that the presence of the $I^2  \Omega^{-2}$ term in the equations
upsets the delicate balance between nonlinearity and dispersion which allowed
for the appearance of localized, and even compact, solitary waves.
Only periodic solutions exist when $I \ne 0$.
To show this, we take the case $\beta_0 \ne 0$, $\beta_1 = 0$, $\lambda = 1$.

When $E \ne 0$, the changes of variable and function \eqref{changes1} give the
following non-dimensional equation,
\begin{equation} \label{I_1}
 \omega^{'2} = - \frac{\omega^2 - 1}{\omega^2}
  \left[ \omega^4 + \left(1 - 4\frac{\rho c^2 - \mu}{\mu b \Omega_0^2}\right)\omega^2
    - 4 \frac{\rho c^2 I^2}{\mu b \Omega_0^6} \right].
\end{equation}
Now recall that the existence of a localized (pulse or kink) solitary
wave is subordinated to the existence of a double root in the right hand-side
of these equations \cite{peyrard}, \cite{sacco04}.
According to the choice for $c$,
Eq.\eqref{I_1} can be written in the form
\begin{equation}
 \omega^{'2} = - \frac{\omega^2 - 1}{\omega^2}
  ( \omega^4 \pm \gamma^2 \omega^2 - \delta^4),
    \quad
\gamma \equiv \left|1 - 4\frac{\rho c^2 - \mu}{\mu b \Omega_0^2}\right|^{1/2},
\quad
\delta \equiv  \left[4 \frac{\rho c^2 I^2}{\mu b \Omega_0^6}\right]^{1/4},
\end{equation}
where the plus sign is taken
when $\rho c^2 \le  1 - 4(\rho c^2 - \mu)/(\mu b \Omega_0^2)$
and the minus sign is taken
when $\rho c^2 \ge  4(\rho c^2 - \mu)/(\mu b \Omega_0^2) - 1$.
A simple analysis of these equations reveals that their right hand-side cannot
have a double root and remain positive at the same time.
It follows that here, \emph{there are no solitary waves, only
periodic waves}.

Similarly when $E = 0$, the changes of variable and function \eqref{changes2} give
an equation which can be rewritten as
\begin{equation} \label{pp2}
 \omega^{'2} =  - \frac{1}{\omega^2}
  ( \omega^6 - \omega^4 + \delta),
    \qquad \text{where} \qquad
\delta \equiv  \rho c^2 \beta_0 I^2 \left[\frac{\mu b}{4(\rho c^2 - \mu)}\right]^2.
\end{equation}
Here the discriminant of the cubic in $\omega^2$ is $-\delta(27 \delta -4)$;
it is zero when $27 \delta = 4$, but the corresponding right hand-side of \eqref{pp2}
is negative.

We conducted the same analysis when $\beta_1 \ne 0$ and checked
that again, the other solitary localized waves (pulses, kinks,
compact-like) disappear due to the introduction of the singular
term $I^2/\Omega^2$. 
A direct and straightforward analysis of
\eqref{1st_integral} makes it clear that this conclusion does not
depend on the particular choice of $\Sigma$ made here. 
When the generalized shear modulus satisfies the empirical inequality $Q>0$,
localized waves are bound to disappear when $I \neq 0$.

\subsection{Pre-stretch}

When the solid is pre-stretched, the principal strain invariant $I_1$ is given by
\eqref{pw6} and equation \eqref{1st_integral} is changed
accordingly.
In the case $I=0$, Eq.\eqref{eqn_pw} is replaced with
\begin{equation}
 \rho c^2 (\beta_0 \Omega^{' 2} + \frac{\beta_1}{4}\Omega^{' 4})
  = \mu \lambda^2 \left[ 1 + \frac{b}{2} (2 \lambda^{-1} + \lambda^2)\right]
                           (\Omega^2 - \Omega_0^2)
    + \mu \frac{b}{4} \lambda^4 (\Omega^4 - \Omega_0^4),
\end{equation}
where $\Omega_0$ is the value of $\Omega$ when $\Omega^{'2} = 0$
(it is related to $E$ through $2 E = \Sigma(\Omega^2_0)$.)

Hence the analysis is hardly modified by  the introduction of pre-stretch.
Methodologically and qualitatively, the results of Sections 4.1 and 4.2 apply here;
quantitatively, it makes no sense to compare the unpre-stretched situation with the
pre-stretched case because $\Omega_0$ and $c$ are arbitrary.

\subsection{Other strain-hardening power-law solids and fourth-order elasticity}

When $n$ in \eqref{11} is an integer other than 2, the analysis is not
overly modified.
For instance, the factorizations occurring at $n=2$ on the right hand-side
of \eqref{non_dimensional1} and \eqref{non_dimensional2} are still in force,
and the bracketed term is now a polynomial of degree $n-1$ in $\omega^2$.
All the results derived at $n=2$ are easily extended.
We leave the case where $n$ is not integer an open question.

As an example, we seek a pulse solitary wave in an unpre-stretched
$n = 3$ power-law dispersive solid.
We find that the counterpart to \eqref{eqn_pw} is
\begin{equation} \label{eqn_pw3}
 (\mu - \rho c^2 + \frac{\mu b}{3}\Omega^2 + \frac{\mu b^2}{27}\Omega^4)\Omega^2
  + \rho c^2(\beta_0 \Omega^{'2}  + \frac{\beta_1}{4} \Omega^{'4}) = 2 E.
\end{equation}
We look for the linearly polarized, pulse solitary wave corresponding to the
case $\beta_0 \ne 0$,  $\beta_1 = 0$,  $E= 0$.
To make a meaningful comparison with the $n=2$ case, we perform the same changes of
variable and function as in \eqref{changes2}.
We find that the counterpart to \eqref{non_dimensional2} is
\begin{equation}
\omega^{' 2} = - \omega^2 \left(\frac{4}{3}\omega^2 
   + \frac{16}{27}\frac{\rho c^2 - \mu}{\mu}\omega^4 - 1 \right).
 \label{non_dimensional5}
\end{equation}
This differential equation can actually be solved using hyperbolic functions.
Rather than present the details of that long resolution, we rapidly discuss the
effect of having a ``stiffer'' power-law material --- in the sense that $n$ goes from 2 to 3
while $\mu$ and $b$ remain the same.
When $n=2$ the maximal amplitude of the wave is $\omega(0) = 1$, see \eqref{non_dimensional2};
when $n=3$ the maximal amplitude is found by solving the bracketed biquadratic on the
right hand-side of \eqref{non_dimensional5}.
An elementary comparison shows that the maximal amplitude at $n=3$ is always larger than $1$.
Hence the ``stiffening'' of the material increases the amplitude of the pulse solitary wave.
Clearly, the same conclusion can be reached for the compact wave corresponding to the case
 $\beta_0 = 0$,  $\beta_1 \ne 0$,  $E= 0$.

For the special case of fourth-order elasticity \eqref{hamil},
we find that 
\begin{equation}  \label{coeff1}
i_2 = - \Omega^2/4, \qquad i_3 = 0,
 \qquad \text{ so that }   \qquad
\Sigma(\Omega^2) = \mu \Omega^2/4 + \nu_4 \Omega^4/16.
\end{equation}
Clearly, by comparing this equation with \eqref{eqn_n_2}, 
and by identifying $\nu_4$ with $2 b$, we find the exact
same results for the fourth-order
elasticity theory of incompressible dispersive solids as
we have for the $n=2$ power-law solid.
Hence in particular, transverse pulses and kinks with compact support 
are possible in a forth-order elastic solid with constitutive 
equations \eqref{hamil} and \eqref{13}.


\section{Concluding remark: \\ A vector MKdV equation}

To conclude, we go back to the general governing equation (\ref{pw10}).
We do not specialize the constitutive relations for $\Sigma$ and $\alpha$,
but we perform a moving frame expansion with the new scales
$s = z-c t$, $\tau = \epsilon t$.

We assume that $W$ is of the form
\begin{equation}
 W = \epsilon^{1/2} w, \quad \text{where} \quad
  w = O(1).
\end{equation}
Then $\Omega = |W| = \epsilon^{1/2}|w|$ and we expand the terms in (\ref{pw10}) as
\begin{align}
& (Q W)_{zz} = \epsilon^{1/2} Q(0) w_{s s}
    + \epsilon^{3/2} Q'(0) \left( |w|^2 w \right)_{s s} + \ldots,
    \notag \\
& \rho W_{tt} = \epsilon^{1/2} \rho c^2 w_{s s} -  2\epsilon^{3/2} \rho c w_{s \tau}  + \ldots,
   \notag \\
& (\alpha W_{t t})_{zz} = \epsilon^{1/2} c^2 \alpha(0) w_{s s s s}
   - 2 \epsilon^{3/2} c \alpha(0) w_{s s s \tau} + \ldots
\end{align}
In order to recover the linear wave speed at the lowest order (here, $\epsilon^{1/2}$)
given by $Q(0) = \rho c^2$, we must assume that
\begin{equation}
Q(0) = O(1), \quad \text{and} \quad
\alpha(0) = O(\epsilon) = \epsilon \alpha_0 \text{ (say)},
\end{equation}
where $\alpha_0$ is a constant of order $O(1)$.

Then we find at the next order that
\begin{equation}
Q'(0) \left( |w|^2 w \right)_{s s} + c^2 \alpha_0 w_{s s s s}
 =   -  2 \rho c w_{s \tau},
\end{equation}
which we integrate once with respect to $s$ to get the vectorial MKdV
equation of Gorbacheva and Ostrosky \cite{Gorba},
\begin{equation}
 w_{\tau} + q\left( |w|^2 w \right)_{s s} + p w_{s s s}
 =   0,
\end{equation}
where here $q \equiv  Q'(0)/(2\rho c)$ and $p \equiv c \alpha_0/(2\rho)$.

In this way we have clarified the status of the Gorbacheva and
Ostrovsky's beautiful results in the framework of the general theory
of dispersive hyperelasticity.


\end{document}